\documentclass[prl, twocolumn, showpacs, superscriptaddress,floatfix]{revtex4}

\usepackage{amsmath,latexsym}
\usepackage{float}
\usepackage{epsfig}

\begin{document}
\title{Strain-Rate Frequency Superposition (SRFS) - A rheological probe of structural relaxation in soft materials}

\author{Hans M. Wyss}
\affiliation{Department of Physics \& HSEAS, Harvard University,
Cambridge, MA 02138}
\author{Kunimasa Miyazaki}
\affiliation{Department of Chemistry,
Columbia University, New York, NY 10027}
\affiliation{The Research Institute of Kochi University of Technology, Tosa Yamada, Kochi 782-8502, Japan}
\author{Johan Mattsson}
\affiliation{Department of Physics \& HSEAS, Harvard University,
Cambridge, MA 02138}
\author{Zhibing Hu}
\affiliation{Department of Physics, University of North Texas, Denton, TX 76203}
\author{David R. Reichman}
\affiliation{Department of Chemistry,
Columbia University, New York, NY 10027}
\author{David A. Weitz}
\affiliation{Department of Physics \& HSEAS, Harvard University,
Cambridge, MA 02138}

\begin{abstract}
\noindent The rheological properties of soft materials often exhibit
surprisingly universal linear and non-linear features. Here we show
that these properties can be unified by considering the effect of
the strain-rate amplitude on the structural relaxation of the
material. We present a new form of oscillatory rheology, Strain-Rate
Frequency Superposition (SRFS), where the strain-rate amplitude is fixed as the
frequency is varied.
We show that SRFS can isolate the response due to structural relaxation,
even when it occurs at frequencies too low to be accessible with standard
techniques.
\end{abstract}
\date{\today}

\maketitle

Concentrated suspensions\,\cite{Mason_glass_rheo_prl},
pastes\,\cite{Cloitre_paste_rheo_PRL_2000},
emulsions\,\cite{Mason_emulsion_rheo_prl},
foams\,\cite{Cohen_foam_rheology}, or associative polymer
systems\,\cite{Hyun2002,Daniel2001} represent commonly encountered
examples of soft solids. Their structures are often metastable, with
dynamics strongly reminescent of that of glasses. They are thus
often termed soft glassy
materials\,\cite{Sollich_softglassy_PRL_1997}. Despite the wide
diversity of these materials, their rheological response, as
characterized by the complex shear modulus
$G^*(\omega)=G'(\omega)+iG''(\omega)$, is remarkably similar. The
storage modulus $G'$ is only very weakly dependent on frequency. The
loss modulus $G''$ also shows a very weak frequency dependence and
is much larger than the response due to the continuous phase fluid.
It often exhibits a shallow minimum near the lowest experimentally
accessible frequencies. At even lower frequencies, there should be a
crossover from solid-like to liquid-like behavior which is signaled
by a pronounced peak in $G''(\omega)$, reflecting the structural
relaxation. Unfortunately, the relaxation frequencies are often much
too low to be accessed by standard linear rheological measurements.
A typical example of this linear viscoelastic behavior is shown in
Fig.\,1(a) for soft hydrogel spheres.
Remarkably, the similarities in the rheological response of these
materials extend even to nonlinear measurements,
characterized by strain-dependent viscoelastic measurements
performed at constant $\omega$, varying the strain amplitude. 
Above a critical yield strain, the storage modulus exhibits a power-law
decay $G'(\gamma_0) \propto {\gamma_0^{-\nu'}}$. 
By contrast, the loss modulus exhibits a well-defined peak,
before falling as a power-law $G''(\gamma_0) \propto
{\gamma_0^{-\nu''}}$; typically with $\nu'' \approx \nu'/2$.
The pronounced peak in
the loss modulus is a remarkably 
robust feature of soft glassy materials
\,\cite{Mason_glass_rheo_prl,Mason_emulsion_rheo_prl,Daniel2001,Hyun2002,Robertson2005}.
The ubiquitousness and similarity of the rheological response, both
linear and nonlinear, of so many soft materials suggests that the
response is governed by a common underlying mechanism. A possible
clue to the origin of this behavior comes from a proposed
explanation for the response of a supercooled
fluid\,\cite{Kuni_paper}. Within this picture, the peak in
$G''(\gamma_0)$ observed in nonlinear measurements is directly
related to a decrease of the structural relaxation time with
increasing shear rate
\,\cite{Sollich_softglassy_PRL_1997,yamamoto1998,berthier2002,fuchs2002,miyazaki2002};
the applied strain drives the relaxation and forces it to
a higher frequency, where it is directly probed. While this picture
provides an excellent description of a colloidal supercooled liquid,
the underlying physical concept should be much more generally
applicable. If this link is verified, the relaxation could be probed 
using the nonlinear response, even 
when the relaxation frequency is experimentally
inaccessible. This would provide a new probe of the dynamics and rheology of soft materials.
However, there have been no attempts to exploit this behavior and
explore its general applicability to a variety of different
materials.

In this Letter, we show that the typical linear and nonlinear
viscoelastic behavior of a wide variety of soft materials can be
explained by a common mechanism
\,\cite{Sollich_softglassy_PRL_1997,yamamoto1998,berthier2002,fuchs2002,miyazaki2002}.
We introduce a new approach to oscillatory rheology;
by keeping the strain rate constant during the measurement we can
directly probe the structural response. The data exhibits a
characteristic scaling, allowing us to clearly isolate and probe the
response due to structural relaxation. This new technique, which we
call Strain-Rate Frequency Superposition (SRFS), should be generally
applicable to the study of many soft materials.

\begin{figure} [floatfix]
\begin{center}
\includegraphics[width=0.45\textwidth]{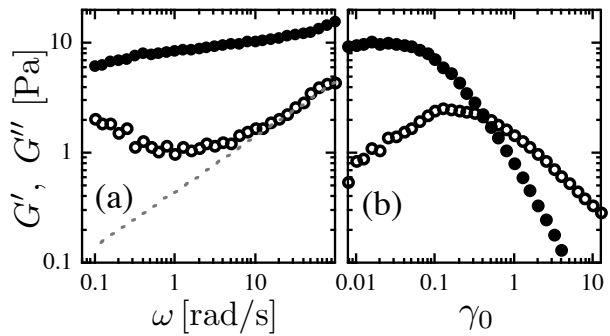}
\end{center}
\caption{ \label{Figure1} Oscillatory measurements
of $G'$(solid circles) and $G''$(open circles) for a suspension
of hydrogel particles. (a)~Linear frequency dependent measurement at
$\gamma_0=0.02$. 
The dashed line is proportional to $\sqrt{\omega}$.  (b)~Strain dependent
nonlinear measurement at $\omega=1\ {\rm rad/s}$.
}
\end{figure}

To elucidate the viscoelastic response, we use a system of
hydrogel microspheres dispersed in water. They consist of
interpenetrating polymer networks of cross-linked poly
N-isopropylacrylamide and polyacrylic acid \cite{Hu:2004,Xia:2004}.
The sample studied has a polymer weight fraction $c_p=0.0186$ and at
low volume fraction the particles have a radius of $R \simeq 95
{\rm nm}$. All rheological measurements are performed with a
strain-controlled rheometer (ARES), applying an
oscillatory strain $\gamma(t)$ to the sample and measuring the
resultant time-dependent stress $\sigma(t)$.

This system of soft spheres shows the behavior typical of soft
materials, both in its linear and nonlinear viscoelastic behavior,
including the ubiquitous peak in $G''(\gamma_0)$ that typefies the
rheology. The measurements shown in Fig.\,1 represent the most
commonly used types of oscillatory rheological measurements. They
depend on two variables only, the frequency $\omega$ and the strain
amplitude ${\gamma}_0$; the applied oscillatory strain for each data
point is ${\gamma(t)}={\gamma}_0  \sin(\omega t)$. Frequency
dependent measurements [Fig.\,1(a)] are usually performed in the
linear viscoelastic regime at a constant $\gamma_0\ll 1$, while
strain dependent measurements [Fig.\,1(b)] are perfomed at constant
$\omega$, probing the viscoelastic response at a fixed time scale
$2\pi/\omega$. However, the probe frequency might not be the only
important time scale. Instead, as the strain rate becomes large, it
can itself drive the slow structural relaxation process at the time
scale of the imposed strain rate. For oscillatory measurements, the
natural scale of strain rate is its amplitude $\dot
\gamma_0={\gamma_0 \omega}$. Thus, if a structural relaxation
process is present, its time scale $\tau(\dot{\gamma_0})$ should
depend on the applied strain rate amplitude,
\begin{equation}
\frac{1}{\tau(\dot{\gamma_0})}  \approx {\frac{1}{\tau_0} + K  \dot{\gamma_0}^\nu}
\end{equation}

However, in traditional strain or frequency dependent rheological
measurements, the amplitude of the strain rate varies for each data
point for both frequency and strain dependent measurements. This
makes it difficult to discern the effects of strain rate on the time
scale $\tau({\dot \gamma_0})$ of a structural relaxation process.

\begin{figure} [floatfix]
\begin{center}
\includegraphics[width=0.5\textwidth]{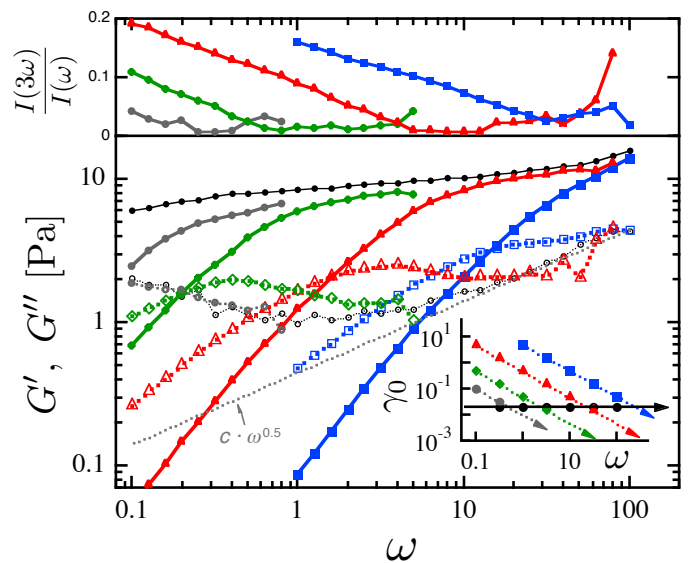}
\end{center}
\caption{ \label{Figure2}  (Color online) Frequency dependent
storage $G'(\omega)$ (solid symbols) and loss modulus $G''(\omega)$
(open symbols), measured in constant rate oscillatory measurements
at different strain rate amplitudes $\dot{\gamma_0}={\gamma_0
\omega}$;  $\dot{\gamma_0}=0.01\
\rm{s}^{-1}$ (grey circles), $0.05\ \rm{s}^{-1}$ (green diamonds),
$0.5\ \rm{s}^{-1}$ (red triangles), $5\ \rm{s}^{-1}$ (blue squares).
The linear oscillatory measurement of $G'$ and $G''$ are shown by the black circles.
Inset:
Strain amplitude $\gamma_0$ versus $\omega$ corresponding
to the constant rate measurements in the main graph.  Top: Corresponding ratios of the third to the first harmonic response $I(3 \omega) /  I(\omega)$ as a function of $\omega$. }
\end{figure}

We circumvent this problem by performing frequency dependent
rheological measurements at constant strain rate amplitude ${\dot
\gamma_0}$. This `constant rate frequency sweep' is achieved by
applying a strain amplitude that is inversely proportional to the
oscillation frequency, as shown in the inset of Fig.\,2. We perform
a series of such measurements with $\dot{\gamma_0}$ ranging
from $0.01\,s^{-1}$ to $5\,s^{-1}$, shown in the main graph of Fig.\,2. For
comparison, the linear viscoelastic measurement (see Fig.\,1)
performed at a constant strain amplitude (circles) is also included.

The viscoelastic response depends strongly on the strain rate
amplitude. With increasing ${\dot \gamma_0}$, the entire frequency
dependent response is shifted towards higher $\omega$, as expected
from Eq.\,(1). Remarkably, however, the general shape of the
frequency dependent storage and loss moduli is surprisingly
insensitive to ${\dot \gamma_0}$. At low frequencies the response is
liquid-like: $G''$ is larger than $G'$ and both moduli scale with
frequency as a power law, with a ratio of $\approx2$ between the corresponding exponents. At higher frequencies there is a pronounced peak in
the loss modulus, followed by a shallow minimum and a final slow
increase at the highest frequencies. The storage modulus rises
continuously until it reaches a plateau where it shows only a very
weak frequency dependence.
The response remains surprisingly harmonic for all our measurements\,\cite{Hyun2002} with a ratio of the third to the first
harmonic component of the response that is always smaller than 20\%; in the relevant frequency range around the peak position this ratio is smaller than 10\%, as shown in the top graph of Fig.\,2.  This justifies our treatment of the data in terms of the viscoelastic moduli $G'$ and $G''$.
\begin{figure} [floatfix]
\begin{center}
\includegraphics[width=0.47\textwidth]{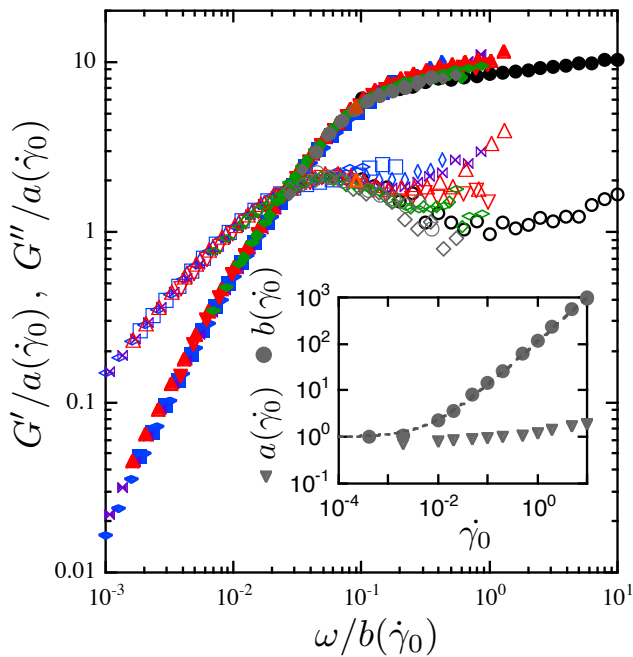}
\end{center}
\caption{ \label{Figure3} Constant rate frequency sweep measurements
shifted onto a single master curve. Inset: Corresponding amplitude
and frequency shift factors $a({\dot \gamma_0})$ and $b({\dot
\gamma_0})$ versus $\dot{\gamma_0}$. }
\end{figure}

To highlight the striking similarities in the shape of the
viscoelastic response, we rescale the moduli in both magnitude and
frequency, using ${G_{\rm scaled}}^*(\omega) = G^*(\omega  / b({\dot
\gamma_0})) / a({\dot \gamma_0}) $ , where $a({\dot \gamma_0})$ and
$b({\dot \gamma_0})$ are the scaling factors for the magnitude and
frequency, respectively. All the data can be superimposed onto a
single master curve, as shown in Fig.\,3. The scaling factors are
plotted in the inset to Fig.\,3 as a function of ${\dot \gamma_0}$.
The value of $a({\dot \gamma_0})$ varies by less than a factor of 3
over the entire range of ${\dot \gamma_0}$ accessed, whereas the
frequency scaling factor $b({\dot \gamma_0})$ depends strongly on
${\dot \gamma_0}$. At large strain rates we observe a power-law
dependence, $b({\dot \gamma_0}) \propto {\dot \gamma_0^\nu}$, with
$\nu \approx 0.9$; at low strain rates it asymptotically approaches
unity, which suggests that in this regime the imposed strain
rate is sufficiently low that it no longer impacts the viscoelastic
behavior. 
The rate dependence of the scaling factors is consistent with the 
behavior expected from Eq.\,(1), with 
$b({\dot \gamma_0}) = {\tau_0 / \tau({\dot \gamma_0})}$.

The observed scaling behavior is a direct consequence of a
structural relaxation process, which leads to a decrease in
$G'(\omega)$ and a concomitant peak in $G''(\omega)$ in linear
viscoelastic measurements. However, this structural relaxation
occurs at too low a frequency to be directly observed in the data in
Fig.\,1(a); nevertheless the incipient behavior is observed in the
linear viscoelasticity at low frequencies and is responsible for the
peak observed in $G''$ in the nonlinear behavior.

The structural relaxation time decreases as $\dot{\gamma_0}$
increases, as expected from Eq.\,(1). The relaxation peaks in the
data shown in Fig.\,3 occur when $2\pi/{\omega}$ becomes comparable
to $\tau(\dot{\gamma_0})$. To isolate this contribution due to
structural relaxation, we must separate the peak from the subsequent
rise in the data at higher frequencies. Closer inspection of the
data reveals systematic differences between the scaled curves at the
highest frequencies. This suggests that a significant  contribution
to the response does not depend on strain rate at sufficiently high
$\omega$. The high frequency viscoelastic response in compressed
emulsions exhibits a $\sqrt{\omega}$-dependence, attributed to
viscous flow along randomly oriented slip planes \cite{Liu:1996};
this results in a contribution that scales as ${\sqrt \omega}$. The
high frequency linear response of our system shows the same
behavior, as shown by the dashed line in Fig.\,1(a). We thus
subtract a component $c  {\sqrt \omega}$ from each of the data sets,
where the constant $c$ is determined from a fit to the high $\omega$ response; 
a single value of $c$ is used for all the data, independent 
of $\dot{\gamma}$.
The resultant loss moduli now
exhibit well defined peaks, as shown in Fig.\,4(a); the component
subtracted is also shown as the dashed line.

The shapes of the resulting relaxation peaks are remarkably similar
at each value of ${\dot \gamma_0}$. To highlight this, we scale all
the data onto a single master curve by rescaling the frequency, as
shown by the uppermost data set in Fig.\,4(b); because we have
subtracted the $\sqrt{\omega}$ component, we do not need to scale
the amplitude. The scaling of the data suggests that the shape of
the relaxation spectrum remains unchanged as the strain rate
amplitude is varied. This should provide important new insight about
the nature of the relaxation and the distribution of time scales
that make up the peak.
\begin{figure} [floatfix]
\begin{center}
\includegraphics[width=0.5\textwidth]{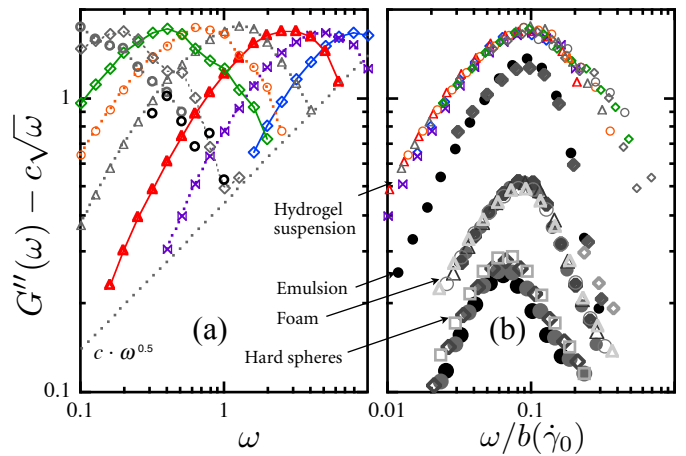}
\end{center}
\caption{ \label{Figure4} (a) Loss modulus $G''$ for a series of
constant strain rate oscillatory measurements with a high frequency
contribution $c  \sqrt{\omega}$ subtracted (dotted line).
(b) The uppermost curve shows the same data shifted onto a single master curve. 
Additional master curves shown: 80\,vol\% oil in water emulsion, aqueous foam 
(Gillette Foamy), PMMA hard sphere suspension at $\phi \approx 0.58$.
Each data set has been scaled for clarity. }
\end{figure}

The observed scaling suggests that $G^*$ can be
expressed as a function of frequency and strain amplitude:
\begin{equation}
G^*(\omega, \gamma_0) \approx G_{\rm R}^*\left({\omega /  b({\gamma_0 \omega}) }\right) + G_{\rm HF}^*\left({\omega}\right) ,
\end{equation}
where $G_{\rm R}^*\left({\omega}\right)$ describes the functional
form of the linear viscoelastic behavior due to the structural
relaxation process, and $G_{\rm HF}^*\left({\omega}\right)$
describes high frequency contributions that are independent of
$\gamma_0$; caution must be used for $\omega<\omega_\mathrm{max}$, where the slowest relaxation must dominate\cite{Wyss_topublish}.
This expression for $G^*$ reflects both the linear and
the incipient nonlinear response of the system. For low enough
values of strain, Eq.\,(2) describes the complex modulus in the
linear viscoelastic regime, with $b({\dot \gamma_0}) \approx 1$. At
larger strains, the structural relaxation shifts towards higher
frequencies, scaling as $b({\dot \gamma_0})$. This allows us to
establish a direct link between linear and nonlinear rheology
through the strain-rate dependence of the structural relaxation
time.  To further illustrate the physical origin of this link, we provide 
a schematic movie in the supplemental information\cite{suppl_info}.

Our results provide an important new measurement technique for
probing the rheological properties of soft materials. By maintaining
a constant strain-rate amplitude during our measurement, we can
isolate the structural relaxation; by varying the amplitude of the
strain rate, we force the frequency of this structural relaxation
into a regime accessible to the rheometer. This allows us to
directly probe this structural relaxation, even if it is at a
frequency too low to be easily accessible. By scaling the data, we
highlight the low-frequency contributions of the structural
relaxation. A wide variety of soft materials exhibit such a
low-frequency structural relaxation; thus, this method should be
very broadly applicable. To illustrate this generality, we perform
similar experiments on several representative samples of soft
materials, and isolate the structural relaxation by scaling the peak
data onto master curves. Similar behavior is observed for all
samples as shown in Fig.\,4(b), where we plot master curves obtained
for a compressed emulsion (80\% silicon oil drops of $\approx
50\,\mu$m dia in water, stabilized by 10\,mM SDS), a foam (Gillette
Foamy), and a hard-sphere colloidal suspension (PMMA in
cycloheptylbromide/decalin). In all cases, good scaling is observed;
moreover, differences in shape are also apparent, reflecting
different contributions to the relaxation spectrum.

The scaling employed here is reminiscent of an approach that
has been successfully employed for polymer melts. Rheological
measurements of $G'(\omega)$ and $G''(\omega)$ for different
temperatures are shifted onto a single master curve that reflects
the viscoelastic behavior in a dramatically extended range of
frequencies. This technique is called time-temperature superposition
(TTS) \cite{Heijboer_TTS_1956,Ferry_TTS_1980}. By analogy, we call
our approach strain-rate  frequency superposition (SRFS). With SRFS,
we can obtain detailed information on the strain rate dependence of
the slow relaxation process. Our picture combines linear and
nonlinear oscillatory rheology into a unified picture, thereby
accounting for the unusual behavior typically observed in many soft
materials. Our experimental results show that nonlinear viscoelastic
measurements contain useful information on the slow relaxation
dynamics of these systems. Their yielding behavior directly probes
the structural relaxation process itself, shifted towards higher
frequencies by an applied strain rate. 
Physically, this suggests that this structural relaxation, driven by
an imposed strain rate, results in the same response as the
equilibrium structural relaxation at a much lower frequency. This is a
surprising result, as for the steady shear case theory\cite{fuchs2002} suggests a rate-dependent
change of the high frequency flank of the relaxation; for the systems studied 
we do not observe this in our oscillatory measurements.
By exploiting this behavior,
SRFS isolates the shape of the structural relaxation,
even if it is too slow to be accessed with conventional oscillatory
rheology. 
Thus, SRFS
will provide new insight into the physical mechanisms that govern
the viscoelastic response of a wide range of soft materials.

This work was supported by the NSF (DMR-0507208,
DMR-0602684, CHE-0134969), the Harvard MRSEC
(DMR-0213805), the Hans Werth\'{e}n, the
Wenner-Gren, and the Knut, and Alice Wallenberg Foundations.

\end{document}